# A Study of Dynamic Lorentz Force Detuning of 650 MHz $\beta_g$ = 0.9 Superconducting Radiofrequency Cavity

Abhay Kumar[1], Arup Ratan Jana[2] and Vinit Kumar[2]

[1]Proton Linac and Superconducting Cavities Division

[2]Materials & Advanced Accelerator Sciences Division

Raja Ramanna Centre for Advanced Technology, Indore – 452013, India

Email: abhay@rrcat.gov.in, abhayk99@gmail.com

**Abstract**

*The small bandwidth of superconducting cavities makes the study of dynamic Lorentz force detuning and its compensation indispensable in case of pulsed mode operation of high gradient accelerators. In this paper, we present the study of this detuning and also propose an optimized design for five cell 650 MHz $\beta_g$ = 0.9 elliptic superconducting cavities, which will be used in the high energy section of the 1 GeV H⁻ linear accelerator for the proposed Indian Spallation Neutron Source project, by suitably inserting the inter-cell stiffeners. The paper presents a sequential design methodology which starts with study of static Lorentz force detuning and tunability; and progresses to find out the structural modes and related dynamic detuning values by performing transient structural dynamics calculations. The developed methodology is general in nature and can be used for a three dimensional model of any geometry. The work will be useful for optimizing the design against dynamic Lorentz force detuning of superconducting radiofrequency cavities of any shape.*

## I. Introduction

Indian Spallation Neutron Source (ISNS) is envisioned to facilitate neutron based multi-disciplinary research in the fields of condensed matter physics, material science, chemistry, biology and engineering sciences. The proposed design of the ISNS consists of a pulsed linear accelerator (LINAC) that delivers 4 mA pulse current, 1 GeV H⁻ ion beam at a pulse repetition rate (PRR) of 50 Hz [1] and a duty cycle (percentage ratio of beam bunch length and the time period) of 10%. ISNS LINAC will use superconducting radio frequency (SRF) cavities to avail several advantages such as less power dissipation on the cavity wall and the possibility of operating at larger beam aperture radius that allows higher beam current to be accelerated [1, 2]. Two sets of multi-cell elliptic cavities will be used – one set for medium energy that will accelerate the H⁻ beam from 200 MeV to around 500 MeV, and the other set for high energy that will accelerate the beam from 500 MeV to 1 GeV [3, 4, 5]. Electromagnetic design of 650 MHz, five cell SRF cavity of the high energy section has been recently optimized for a particle velocity of 0.9 times that of light ($\beta_g = 0.9$) [1].

In an RF cavity, electromagnetic field induces surface current and surface charges on the wall of the cavity. This in turn, generates a Lorentz pressure on the cavity surface which, depending on the RF pulse repetition rate, sets up mechanical vibrations in the structure. As a consequence, the resonating frequency of the cavity starts changing dynamically. This detuning is called dynamic Lorentz force detuning (LFD) [6].

The detuning changes the impedance of the cavity and if it is not compensated swiftly, then the cavity would reflect a fraction of the input power due to loss of impedance matching with the power source [6]. The half power bandwidth is inversely proportional to the quality factor; thus the detuning tolerance of the SRF cavities is extremely limited [7].

In order to reduce the magnitude of the dynamic LFD; the SRF cavities of high gradient pulsed accelerators are stiffened by placing ring stiffeners [8] between the cells as well as between the end cell and the end cover of a vessel that acts as a 2 K liquid helium bath for the niobium cavity. This vessel is made of titanium and is generally known as helium vessel.

We performed a literature survey of the methodologies used in dynamic LFD studies of SRF cavities. Mitchel *et. al.* [8] performed the dynamic LFD studies for SRF cavities of Spallation Neutron Source (SNS) using the computer codes ABAQUS™ for calculations of structural deformations, and Poisson Superfish [10] for calculation of electromagnetic modes and fields. Lin *et. al.* showed that ANSYS™ can be used as a tool for the coupled field analysis for the design of RF cavities [9]. Later, the same tool was used by them to perform coupled field simulation of 500 MHz SRF cavity to calculate frequency shift due to cool-down using MSC/PATRAN™ as the pre-processor to build the finite element model [10]. Losito *et. al.* also showed that ANSYS™ can be used to perform coupled field simulation including Lorentz force detuning studies [11]. They applied this analysis tool on a 352 MHz $\beta_g = 0.7$ four cell SRF cavity and calculated static Lorentz force detuning.. Later, Sun An [12] developed a simulation method by combining the strengths of Poisson Superfish in electromagnetic domain and ANSYS™ in the structural domain and showed it to be a precise tool for simulating changes in the cavity frequency and field flatness due to the elastic deformations by performing measurements on the cavity. This procedure was implemented to perform dynamic Lorentz force detuning studies of 700 MHz, $β_g$ = 0.42 elliptic superconducting RF cavity for the LINAC of Proton Engineering Frontier Project (PEFP) [13]. Bandyopadhyay *et. al.* [14] reported the use of ANSYS™ in electromagnetic, thermal and structural analysis of LINAC cavity. Recently, the elliptic SRF cavity of Cornell Energy Recovery LINAC was optimized for stiffening for minimizing the sensitivity against helium pressure fluctuation using ANSYS™ by performing elastic static displacement calculations in the structural module and resulting change in the electromagnetic mode in the high frequency electromagnetic module [15]. Static LFD calculations were also performed where the Lorentz pressure used in the calculation of structural deformation was calculated using the electromagnetic fields in ANSYS™ [15]. We have also preferred a single software environment of ANSYS™ with its seamless programming interface called ANSYS™ parametric design language (APDL). We generated a fully compatible finite element mesh at the interface of structural mesh in niobium and electromagnetic mesh in vacuum for ensuring one-to-one correspondence between the calculated electromagnetic fields and applied Lorentz pressure on the structural elements. This also ensured an accurate data exchange at every time step of analyses in the high frequency electromagnetic and transient dynamics modules.

The boundary stiffness i.e. the stiffness of the helium vessel at the extremities of the cavity plays a very important role in reducing the Lorentz force detuning [1]. The choice of stiffener location was optimized by varying the boundary stiffness during the design studies of elliptic SRF cavities of SNS [8]. In addition, the choice of two stiffener rings was considered in the design of SNS cavities and also in PEFP cavity [8, 12]. We started with the study of the time domain and frequency domain variation of the Lorentz pressure on the cavity wall. This knowledge was augmented by the knowledge of mechanical modes of the cavity structure at different stiffener locations and was very helpful in allowing us to work with single stiffener design and fixing the helium vessel design prior to the study of the dynamic response.

This paper presents a sequential approach for optimizing the stiffener location for SRF cavities of high gradient pulsed accelerator while quantitatively discussing the significance of constraints that lead to the selection. First, the temporal variation of the Lorentz pressure during a pulse and its Fourier series expansion are obtained. Then, the finite element model of the cavity is discussed. After this we discuss the limit imposed on the choice of stiffener locations from the considerations of feasibility of tuning the cavity and the field flatness. Finally, the results are presented, discussed and concluded with a selection of stiffener location. The work pertains to 650 MHz $β_g$ = 0.9 elliptic SRF cavity of high energy section of ISNS. The procedure developed here is general in nature and can be applied to the medium beta cavities as well as to the low beta non-elliptic cavities.

## II. DYNAMIC LORENTZ FORCE DETUNING

The Lorentz pressure $P$ on the cavity's internal surface is given by [2, 6, 8],

$$P = \frac{1}{4}(\varepsilon_0 E^2 - \mu_0 H^2). \tag{1}$$

$E$ and $H$ are electric field amplitude and magnetic field intensity amplitude respectively.

The Lorentz pressure acts outward near the equator and inward near the iris (Fig. 1). It can be seen that outward pressure of lower magnitude acts over a larger area in the equator region whereas inward pressure of larger magnitude acts over a smaller area in the iris region. If gradually applied, this distribution will reduce the length of each cell of the cavity due to larger moment produced by the forces in the iris region about the equator of the cavity.

The frequency change is proportional to the square of electromagnetic fields. The relation [6, 8] between detuning $\Delta f$ and accelerating gradient $E_{acc}$ is given by:

$$\Delta f = -K_L E_{acc}^2 \tag{2}$$

$K_L$ used in (2) is known as Lorentz force detuning coefficient and has the unit of Hz/(MV/m)$^2$ and the accelerating gradient is 18.6 MV/m for the cavity.

Equation (1) shows that the Lorentz pressure on the cavity walls is proportional to the square of the electromagnetic field amplitude. Therefore, Lorentz pressure pulse shape can be readily derived from cavity voltage pulse shape. The ratio of the dynamic displacement amplitude and the static displacement, known as dynamic amplification factor, depends on the frequency ratio of harmonics of excitation force and the structural mode frequencies. The dynamic amplification factor is larger than unity if the frequency ratio lies between 0 and $\sqrt{2}$ for a lightly damped system [16]. This broad range indicates a possibility of a dynamic amplification of the static Lorentz force detuning. The dynamic amplification factor is less than unity if the frequency ratio is larger than $\sqrt{2}$ [16]. The magnitude of dynamic LFD depends on many parameters like frequency ratio, damping and pulse time structure apart from the accelerating gradient; therefore we have not represented dynamic LFD as a function of the accelerating gradient. While comparing static and dynamic LFD magnitudes, it appears more appropriate to show their magnitudes rather than the corresponding value of $K_L$.

### III. LORENTZ PRESSURE PULSE SHAPE

A typical temporal shape of the cavity voltage due to input RF power during a pulse is shown in figure-2. H⁻ ion beam is injected in the flat top duration of 2 ms. The rising and falling part of the RF pulse shape is determined by the loaded quality factor of the cavity *i.e.* $Q_L$(~ 5 x 10⁶) [1].

It is evident that the Lorentz pressure is periodic and will repeat with the same frequency as that of the input *RF* pulse. The mechanical vibration of the cavity wall in response to the time-dependent Lorentz pressure *P* can be analyzed in terms of its natural modes of vibration. Let the cavity surface be parameterized in terms of appropriately defined variables *u* and *v*. In this notation, the cavity surface has parametric representation given by $x = x(u,v)$, $y = y(u,v)$ and $z = z(u,v)$ and displacement at different points on the cavity wall surface due to vibration is denoted by $\chi(u,v,t)$. Decomposing into Eigen modes of vibration, we can write $\chi(u, v, t) = \sum_m \phi_m(u, v)\eta_m(t)$, where $\phi_m(u, v)$ denote the *m*ᵗʰ orthonormal Eigen mode shape function and $\eta_m$ denotes the amplitude of the *m*ᵗʰ mode. The orthonormality condition for the Eigen mode shape functions is given by

$$\iint \phi_m(u, v)\phi_n(u, v)dS = S\, \delta_{mn}, \qquad (3)$$

where the integration is carried over the surface of the cavity, *S* is the surface area, and $\delta_{mn} = 1$ if *m* = *n*, and $\delta_{mn} = 0$ if $m \neq n$.

As shown in Appendix A, the equation for the evolution of $\eta_m$ is like the forced simple harmonic damped oscillator equation, and is given by

$$\ddot{\eta}_m(t) + 2\xi_m \dot{\eta}_m(t) + \omega_{0m}^2 \eta_m(t) = \frac{1}{M}\iint P(u, v, t)\phi_m(u, v)dS, \qquad (4)$$

where the integration is carried over the surface of the cavity, '·' and '··' denote the first and second derivative respectively, $\omega_{om}$ is the *m*ᵗʰ normal mode frequency of the structure and *M* is the mass of the cavity. Here, $\xi_m$ is the damping for the *m*ᵗʰ mode, expressed as a fraction of critical damping [16]. The space dependent and time dependent parts of the pressure function *P(u,v,t)* are separable, and therefore Eq. (3) can be simplified by expressing *P(u,v,t)* as

$$P(u, v, t) = F(t)\wp(u, v) = \frac{F(t)}{S}\sum_n A_n\, \phi_n(u, v), \qquad (5)$$

where the space dependent part of the pressure function, which is decided by the particular electromagnetic mode excited in the cavity, has been expanded in terms of the Eigen modes of mechanical vibration. The temporal dependence is present in the function $F(t)$, and is described by the pulse shape of RF power inside the cavity. Substituting Eq. (5) in Eq. (4), we get

$$\ddot{\eta}_m(t) + 2\xi_m \dot{\eta}_m(t) + \omega_{0m}^2 \eta_m(t) = F(t) A_m \qquad (6)$$

Since $F(t)$ is a periodic function, we can express it as a Fourier series and write $F(t) = \sum_n f_n \sin(n\omega t + \theta_n)$, where $\omega = 2\pi$ times the repetition rate of the pulses. This model ensures that Eq. (6) is converted into simple homogeneous equation for all other frequencies except $n\omega$. The forced oscillation will appear only at the structural mode frequencies. It also indicates that resonance will occur if $\omega_{0m} = n\omega$ is satisfied for a particular $m$ and $n$. The discrete participating frequencies are the integral multiples of the pulse repetition rate of 50 Hz. Fig. 3 shows the frequency domain representation of the forcing function. It is seen that the structural mode frequencies higher than 250 Hz have progressively lower contributions in the total response. Based on this analysis, an important conclusion can be drawn that there is a radical reduction in amplitudes corresponding to the higher participating frequencies if the structural mode frequencies are higher than 250 Hz.

### IV. THE FINITE ELEMENT MODEL

The internal geometry of the five cell elliptic cavity has been optimized to obtain the maximum value of allowed acceleration gradient, keeping the resonant frequency for the TM010-π mode fixed at 650 MHz [1] and is presented in figure-4. The wall thickness of the cavity has been kept at 4 mm on the basis of studies performed on a similar structure [17].

It is evident from equation (4) that the Lorentz pressure can excite only those structural modes in the cavity, which have their shape function similar to the Lorentz pressure distribution on the surface. Therefore, a 5° sector in azimuthal direction with full cavity length was modeled for transient analysis to capture all longitudinal modes. The stiffeners were modeled with 3 mm thickness of niobium. The material properties used in the calculations have been taken from Ref. 21. The longitudinal modes of vibration depend on the boundary stiffness of the structure. Mitchell *et. al.* [8] modeled the boundary stiffness by springs to simulate the helium vessel. This boundary stiffness can be split into two parts – the bending stiffness of the end cover of the helium vessel and the membrane stiffness of cylindrical portion. Out of this, the former is much smaller than the later and would directly affect the mechanical mode frequency. In our model, we have modeled the full helium vessel without its tuner and joined with the cavity near iris (fig.5); thus the bending stiffness of the end cover is correctly represented. The mode shapes of the excited longitudinal modes show that the cylindrical portion remains stationary and the mode shapes are generated due to bending of the end cover alone, therefore the absence of tuner would have least effect on the final results.

The helium vessel is a cylinder with internal diameter of 504 mm. Its end cover is a combination of torus and flat geometries. It was found in a previous study that the increase in the stiffness of this SRF cavity along with its helium vessel decreases the static LFD [1]. Therefore, the helium vessel stiffness was increased by increasing its wall thickness from 4 mm to 5 mm [18] and increasing the internal torus radius of the end cover from 35 mm to 120 mm. The major contribution in increase of stiffness came from the higher curvature of the end cover; thus this change increased the stiffness appreciably with some decrease in the mass for the same thickness. The frequency of natural modes of vibration is proportional to the square-root of ratio of stiffness and mass [16]; therefore this would shift them upwards. This would make a favorable condition for lowering the dynamic response as the relative importance of higher harmonics of the Lorentz pressure pulse is low.

The welding of the titanium end cover of helium vessel with niobium beam pipe requires a transition piece of 55Ti-45Nb in order to facilitate joining [19]. The 55Ti-45Nb transition piece has been limited to the flat portion of the end cover to avoid metal-forming of 55Ti-45Nb. Therefore, the flat part of the end cover has to be split into two parts. The one joined with cavity beam pipe will be made of niobium and will be a part of cavity end group. The second flat part will be made of 55Ti-45Nb and this will be welded to the niobium part at one end and torus part of the helium vessel end cover at the other end (fig.5). In order to restrict the loss of stiffness arising from a lower elastic modulus of 55Ti-45Nb, the width of the annular plate was kept smallest minimum necessary value allowed on the basis of manufacturing considerations.

ANSYS™ uses Vector Finite Element Method (VFEM) for electromagnetic computations to enforce continuity of the electric field on the inter-element boundaries to ensure accuracy [20]. The mesh density was kept fine near the surfaces and on the axis to ensure accurate calculations of surface fields and fields on the axis. Second order hexahedral elements were used in both the analysis domains.

ANSYS™ parametric design language (APDL) was used to calculate the Lorentz pressure using equation (1) after scaling the normalized surface electromagnetic fields from an electromagnetic modal analysis. The scaling factor was obtained on the basis of the accelerating gradient. This pressure was subsequently applied in the structural analysis for obtaining the deformed geometry of the cavity from which the changed frequency of electromagnetic mode was obtained.

Rayleigh proportional damping model, in which damping is assumed to be a linear combination of the mass and stiffness matrices has been used during the calculations. Following equation gives the damping matrix $C$ in terms of mass matrix $\mathcal{M}$ and stiffness matrix $K$ [22, 23]:

$$C = \alpha \mathcal{M} + \beta K. \tag{7}$$

Here, α is mass damping coefficient and β is stiffness damping coefficient and they are given by:

$$\alpha = \frac{2\xi \omega_m \omega_n}{\omega_m + \omega_n} \quad , \quad \beta = \frac{2\xi}{\omega_m + \omega_n} \tag{8}$$

The amount of damping present in the system for all the modes has been kept at 0.3% of critical damping which is an experimentally determined value during the development of similar cavities [8]. The critical damping corresponds to the smallest value of damping coefficient above which the transient response just attains an exponentially decaying non-oscillatory state [16]. Here, $\omega_m$ has been taken as the first mode that has maximum contribution in transient displacements and $\omega_n$ has been taken as the last contributing mode [22].

In order to get steady amplitude during the RF pulse, the structural transient analysis was run for 4 seconds (200 pulses) by mode superposition method. The time step size for dynamic analyses was kept at 50 μs which is less than one twentieth of the time period of the largest frequency of natural modes of vibration that may get excited during the pulse [24].The electromagnetic resonating frequency and corresponding dynamic LFD values are calculated at every time step of the 200$^{th}$ pulse by using the deformed geometry at that instant.

V. EFFECTS OF TUNABILITY AND FIELD FLATNESS

Figure-24 and 25 of Ref. 1 show that the static LFD decreases as the stiffener position moves radially outward but tuning the cavity becomes impossible if the stiffeners are placed towards the equator. The stiffeners can be positioned without having a substantial effect on the tunability of the cavity (fig.6) [1] if they are placed within the common tangent region of elliptic curvatures of individual cells. The stiffener location can be varied in this region while optimizing against the dynamic effects.

The constraints for restricting the radial position of stiffener also come from field flatness [25] defined as

$$\eta = \left(1 - \frac{E_{max} - E_{min}}{\frac{1}{N}\sum_{1}^{N}(E_i)}\right), \quad (9)$$

Here, $E_i$ is the maximum electric field amplitude in ith cell and $E_{max}$ and $E_{min}$ denote the maximum and minimum of these amplitudes respectively and $N$ represents the number of cells in a cavity. A value of $\eta$ close to 100% is desired to maintain a good synchronism of the accelerating particle with the electromagnetic field.

The slow tuner (mechanical tuner) is operated to tune the cavity after its integration with the helium vessel. The stiffener location should be such that the field flatness remains largely unaffected from these tuning operations. $\eta \geq 98\%$ has been suggested as the design goal in the electromagnetic design of the end cell [26]. The requirement for SNS cavity field flatness was set to be more than 92% [27]. The achieved field flatness in the electromagnetic design of this cavity is 99.4% [1]. Therefore, the stiffener location that ensures a reasonable axial electrical field flatness of 98% when the slow tuner is moved by 1 mm, (i.e. a tuning of ~100 kHz) could be a safe and tolerant design choice.

The field flatness considerations favor the stiffeners between helium vessel and end-half-cells and that between the mid-half-cells at the same radial position as the asymmetric placement of stiffeners produces difference between the change in the optimized length of the end cell and that of the mid cells at the time of tuning which may lead to the loss of field flatness. It was also observed that even a uniform radial location for all the stiffener rings higher than a mean radius of 122.5 mm deteriorates the field flatness value to below 98.2 % when the slow tuner is displaced by 1 mm (fig.7). The consequence of asymmetric placement of stiffeners in the end cell and mid-cell will be discussed later in Section VI

## VI. RESULTS

The static LFD without a stiffener ring is -1.3 kHz. This value is too high for the design of an SRF cavity for a pulsed accelerator and hence, it is evident that the stiffeners are mandatorily required. Table-2 summarizes the results.

The structural modes are well above 250 Hz for the stiffeners' radial position above 116.5 mm and stability is observed in dynamic LFD values. This confirms our earlier prediction based on Fourier expansion of the Lorentz pressure pulse. For the stiffener's radial position at 122.5 mm, we can observe that although the first two structural modes are close to multiples of 50 Hz, the dynamic LFD has not shot up. Table-2 shows that the dynamic LFD is not only small but the location is also less sensitive if the stiffeners' radial location is at a mean radius of 119.5 mm.

Stiffener location at mean radii of 80.5 mm, 101.5 mm and 113.5 mm show resonance and the dynamic LFD figures show a jump. Very high resonance is observed for stiffener mean radius at 80.5 mm as the first natural mode of vibration is at 150 Hz which coincides with Lorentz pressure harmonics having a high coefficient.

The dynamic LFD during a time period after attainment of steady state is shown in fig. 8.

The important figure of merits for mean stiffener radius of 119.5 mm is given in Table 3.

The dynamic LFD is a result of transient displacements of the entire cavity structure which includes shape deformations and length shortening [15]. We found that it is the second one which is more important. If we plot normalized change in cavity length along with normalized dynamic LFD (fig.9), then the association of change in cavity length and the corresponding dynamic LFD becomes clear. Therefore, in order to estimate the dynamic LFD, a plot of transient displacement of end of cavity beam pipe is enough. The scale factor between change in cavity length and the dynamic LFD is 170±5 Hz/µm for all stiffener locations. This scale factor and the cavity end sensitivity are not identical as the displacement of piezoelectric transducer (PZT) produces a different deformation pattern as compared to that produced by the Lorentz pressure.

The transient displacements of the cavity extremities indicate the expected LFD variation during the pulse. Figures 10 to 13 show the change in half of the cavity length for four cases – stiffener at the mean radius of 119.5 mm and at mean radii of 113.5, 101.5 and 80.5 for showing resonant coupling. Since the time-averaged Lorentz pressure always produces an effective compression in the cavity length; the displacements shown in the figures are asymmetric.

### VII. DISCUSSION & CONCLUSION

The core of the ISNS project i.e. the 1 GeV H- linear accelerator will be a pulsed machine. Therefore, the pulsed Lorentz pressure, repeating at a 50 Hz frequency, will result in a time varying oscillations in the cavity structure. The consequence is the dynamic LFD of the cavity. Since the SRF cavities have high quality factor and thus an extremely narrow bandwidth (~120 Hz); this necessitates that this detuning is kept small and manageable by the movement of PZT [6]. It may be noted that the power requirement increases by 6.25% if the uncompensated detuning is half of the bandwidth [6].

The static LFD study gave us an idea of the stiffener position from the consideration of tunability. Thereafter, a systematic and sequential approach towards the estimation of dynamic LFD and stiffener position has been developed. First, we performed the Fourier analysis of the periodic pressure pulses to obtain excitation force in terms of discrete participating frequencies and their corresponding relative amplitudes. The important conclusion of this step is that there is a radical reduction in amplitudes corresponding to the higher participating frequencies if the structural mode frequencies are higher than 250 Hz. Therefore, the system should be modified by changing the radial location of the stiffener rings in such a way that the participating structural mode frequencies are kept above 250 Hz. In addition to this, the resonances should be avoided as the dynamic amplification for a lightly damped system could still be high.

The next step is to calculate the participating structural modes. This can be calculated in two ways – first from the knowledge that the transient excitations coming from the application of Lorentz pressure can produce longitudinally symmetric modes only or by extracting all the modes in a full model and then performing mode superposition analysis by selecting only one mode at a time. The relative magnitude of displacements with respect to the displacements calculated with a large number of modes gives an idea of the participating modes. The first one is an efficient choice as it allows a smaller model to be used in the analyses.

Then, we calculated the transient displacements for 200 RF pulses to get the stabilized transient response using the mode superposition method. The computed displacements during the last pulse were taken for dynamic LFD computation. Since, the process requires several exchanges between the structural and electromagnetic modules; ANSYS™'s programming interface proved to be very useful for the analyses.

The radial position of 119.5 mm is selected for the stiffener ring. The dynamic LFD value is very close to static LFD value for this position as the first structural mode is well above 250 Hz and is also away from resonances. Secondly, the dynamic LFD value appears to be stable with respect to slight change in the position of stiffener ring if a downward position tolerance is applied during manufacturing. Therefore, the selection has the ability to tolerate the difference between the actual fabricated assembly and the FE model arising from the manufacturing errors. It also satisfies all other constraints like field flatness and tunability. Operating the cavity at lower pulse repetition rate during the trials and commissioning stages will be a favorable condition for the dynamic LFD as the frequency ratios of excitation force harmonics and mechanical mode frequencies will become smaller. In addition, the normalized amplitude of harmonics that are near the mechanical modes will be very small.

Stiffener placement at a mean radius of 110.5 mm is not selected in spite of a lower magnitude of dynamic LFD as high values are observed on both the sides of this location. It can be observed that the dynamic LFD is 11480 Hz if the stiffeners are placed at a mean radial position of 80.5 mm. This was the clear signature of a dynamic resonance with a participating frequency of the excitation force that has a high value of normalized amplitude.

An asymmetric stiffener placement i.e. different radial position of the end-cell stiffener as compared to that of the mid-cell stiffener was also considered. Asymmetric placement of stiffeners produces large difference between the change in the optimized length of the end cell and that of the mid cells at the time of tuning leading to the loss of field flatness. A higher radial location of end cell stiffener as compared to that of the mid cell stiffeners reduce the required displacement for compensation but also deteriorate the field flatness substantially. For example, the field flatness deteriorates to 95% when the cavity is extended by 1 mm for tuning for radial location of end cell stiffener at 121 mm and that of the mid cell stiffeners at 91 mm.

This should be noted that we need to compensate for the dynamic LFD only during the flat top of the Lorentz pressure pulse shape. Therefore, a pre-elongation of cavity using the slow tuner can be used to bring down the range of PZT. When a limited pre-elongation of slow tuner is applied such that the incident power is not significantly reflected (~ 1.6 micrometer), the partially compensated dynamic LFD, reduces to 454 Hz only. The tuning force required will be less than 60 N for the PZT and the total range of PZT will be less than 4 micrometer [28]. PZT operated at cryogenic temperatures can provide several tens of Newton combined with a fast response (<100 μs) [6]; therefore this is not a concern.

On the basis of the design study, following design schematic is proposed:

The study revealed the design bases of selection of stiffener position of elliptic multi-cell SRF cavities of high gradient pulsed accelerators. The design approach is sequential and can be used to arrive at a design for any SRF cavity. The limitations coming from electromagnetic design of the cavity and PZT have been adequately addressed. Since the design is an iterative process, changes are inevitable and the approach developed here will be useful for swiftly evaluating the effect of the changes. The Lorentz pressure calculation subroutine is general in nature and can be used for a three dimensional model of any geometry. The work will be directly useful for designing other SRF cavities.

**ACKNOWLEDGEMENTS**

We acknowledge the encouragement received from P D Gupta and S B Roy for publishing this work. We thank Vivek Bhatnagar for discussing the fabrication related constraints. Vivek Bhasin of Bhabha Atomic Research Centre, Mumbai participated in discussions on the choice of appropriate time history solution method and structural damping. Vikas Jain provided useful suggestions for improving the manuscript.

**APPENDIX A**

Here we briefly discuss the proof of Eq. (4). We start with the equation for free oscillation of the cavity surface given in Ref. 16:

$$\ddot{\eta}_m(t) + 2\xi_m \omega_{0m} \dot{\eta}_m(t) + \omega_{0m}^2 \eta_m(t) = 0. \tag{A1}$$

Multiplying Eq. (A1) with $\phi_m(u, v)$ and summing over all m gives

$$\ddot{\chi} = -2\sum_m \xi_m \omega_{0m} \phi_m(u, v) \dot{\eta}_m(t) - \sum_m \omega_{0m}^2 \phi_m(u, v) \eta_m(t). \tag{A2}$$

Physically, the first term of right hand side is damping force term, and the second term is restoring force term. Eq. (A2) is valid for an element of area *dS* at the location (*u,v*) on the surface in the absence of Lorentz pressure. Each term on the right side is the force on element *dS* divided by its mass.

Now, we add the external force responsible for the forced vibration. On the right hand side, it should only make one change – *i.e.*, add the term $[P(u, v, t)dS]/[m(u, v)dS]$, where $m(u, v)$ is mass per unit area. This gives us

$$\ddot{\chi} = -2\sum_m \xi_m \omega_{0m} \phi_m(u, v) \dot{\eta}_m(t) - \sum_m \omega_{0m}^2 \phi_m(u, v) \eta_m(t) + \frac{P(u,v,t)}{m(u,v)}. \tag{A3}$$

Now, multiplying the above equation with $\phi_n(u, v)$, integrating over entire area and using the orthonormality condition, we get

$$\int \sum_m \phi_m(u,v)\ddot{\eta}_m(t)\phi_n(u,v)dS =$$

$$-2\int \sum_m \xi_m \omega_{0m}\phi_m(u,v)\dot{\eta}_m(t)\phi_n(u,v)dS - \sum_m \omega_{om}^2 \phi_m(u,v)\eta_m(t)\phi_n(u,v)dS + \int \frac{P(u,v,t)}{m(u,v)}\phi_n(u,v)dS,$$

(A4)

where $S$ is the total area of the cavity surface. If mass per unit area is constant, and it is assumed that due to stretching of the cavity due to Lorentz pressure, this is not getting affected, we can take $m(u,v)$ outside the integration and we get Eq. (4)

**Tables:**

Table-1 – Values of Internal Geometry Parameters [1]

| Parameters | Value for Mid-Cells | Value for End Half-Cells |
|---|---|---|
| $R_{iris}$ | 50.00 mm | 50.00 mm |
| $R_{eq}$ | 199.925 mm | 199.925 mm |
| $L$ | 103.774 mm | 105.8 mm |
| $A$ | 83.275 mm | 83.275 mm |
| $B$ | 84.00 mm | 84.00 mm |
| $a$ | 16.788 mm | 16.788 mm |
| $b$ | 29.453 mm | 29.453 mm |
| $α$ | 84° | 82.75° |

**Table-2**: Summary of Transient Response of the Cavity

| # | Stiffener Mean Radius (mm) | Frequency of Longitudinally Symmetric Modes (Hz) | Static LFD (Hz) | Dynamic LFD (Hz) | Cavity Stiffness (kN/mm) |
|---|---|---|---|---|---|
| 1 | 122.5 | 296, 551, 757, 882, 925 | -660 | -794 | 17.84 |
| 2 | 119.5 | 280, 534, 745, 787, 932 | -654 | -652 | 15.67 |
| 3 | 116.5 | 266, 515, 733, 877, 939 | -649 | -656 | 13.85 |
| 4 | 113.5 | 252, 493, 721, 876, 943 | -647 | **-1229** | 12.27 |
| 5 | 110.5 | 239, 471, 711, 874, 945 | -648 | -648 | 10.89 |
| 6 | 107.5 | 227, 449, 701, 869, 942 | -653 | -923 | 9.70 |
| 7 | 104.5 | 216, 427, 691, 862, 937 | -659 | -800 | 8.68 |
| 8 | 101.5 | 205, 407, 680, 853, 929 | -669 | **-1120** | 7.78 |
| 9 | 98.5 | 195, 387, 669, 843, 918 | -682 | -670 | 7.01 |
| 10 | 95.5 | 186, 369, 657, 832, 906 | -699 | -877 | 6.34 |
| 11 | 92.5 | 178, 352, 645, 820, 892 | -719 | -815 | 5.75 |
| 12 | 89.5 | 170, 336, 633, 807, 878 | -742 | -916 | 5.23 |
| 13 | 86.5 | 163, 322, 621, 795, 864 | -768 | -1098 | 4.78 |
| 14 | 83.5 | 156, 308, 609, 783, 850 | -796 | -1574 | 4.39 |
| 15 | 80.5 | 150, 296, 597, 771, 836 | -828 | **-11480** | 4.04 |
| 16 | 77.5 | 144, 284, 586, 760, 823 | -862 | -1177 | 3.73 |
| 17 | 74.5 | 139, 273, 575, 749, 810 | -898 | -781 | 3.45 |

Table-3: Figures of merit of design with Mean Radial Position of Stiffener at 119.5 mm

| # | Figures of Merit | Value |
|---|---|---|
| 1 | Required total tuning range for dynamic LFD compensation | 5.3 μm |
| 2 | Field flatness after 1 mm movement of tuner | 98.58% |
| 3 | Cavity end sensitivity while tuning ($\frac{\Delta f}{\Delta \text{cavity}}$) | 208.29 kHz/mm |
| 4 | Tuning efficiency ($\frac{\Delta f}{\Delta \text{tuner}}$) | 123.83 kHz/mm |

**Figure captions:**

Fig.1 – Typical Lorentz pressure distribution on cavity wall

Fig.2 – The normalized amplitude of the Lorentz Pressure Pulse (in black) and the input RF pulse (in blue) as a function of time are shown by the black line and blue line respectively.

Fig.3 – Fourier spectrum of the normalized Lorentz Pressure Pulse Shape for calculation of dynamic LFD; $|F(\omega)|_{Norm}$ is the normalized amplitudes correspond to the participating frequencies.

Fig.4 – The internal geometry of cavity [1]

Fig.5 – The model of cavity with helium vessel

Fig.6 – Tuned LFD when stiffener position is varied for a fixed tuning displacement [1].

Fig.7 – Field flatness reduction with increase in stiffeners' radial position

Fig.8 – Dynamic LFD with Mean Radial Position of Stiffener at 119.5 mm

Fig.9 – Normalized dynamic LFD and change in cavity length with Stiffener's Mean Radial Position at 119.5 mm during a single pulse

Fig.10 – Longitudinal Displacements with Mean Radial Position of Stiffeners at 119.5 mm

Fig.11 – Resonant conditions with Mean Radial Position of Stiffeners at 113.5 mm

Fig.12 – Resonant Conditions with Mean Radial Position of Stiffeners at 101.5 mm

Fig.13 – Resonant Conditions with Mean Radial Position of Stiffeners at 80.5 mm

Fig.14 – Proposed design

**Figures:**

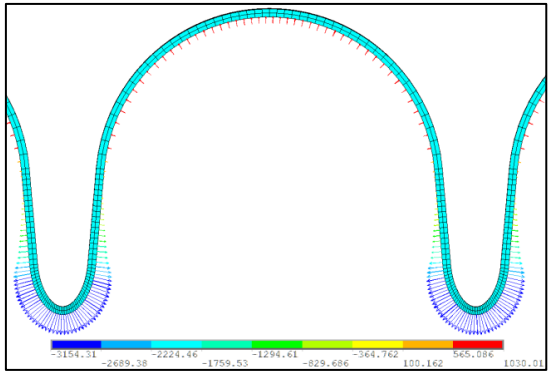 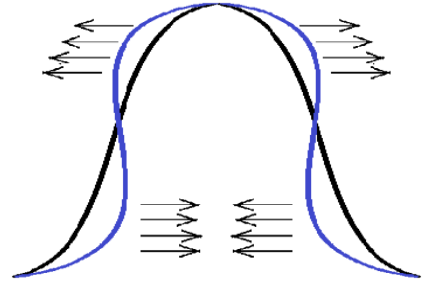

Fig.1

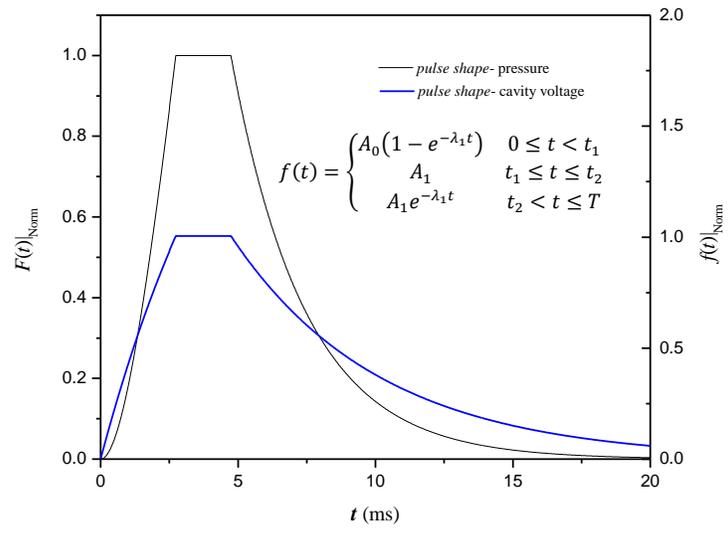

Fig.2

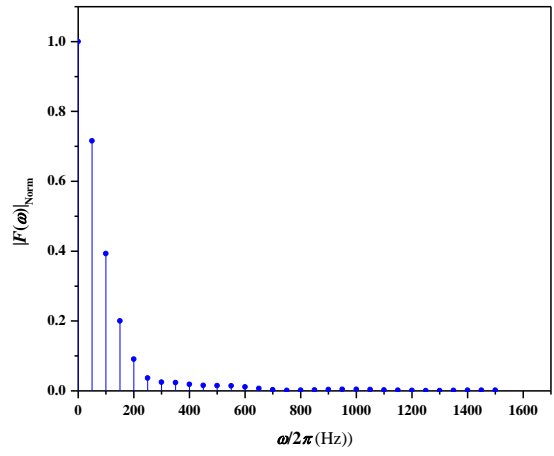

Fig.3

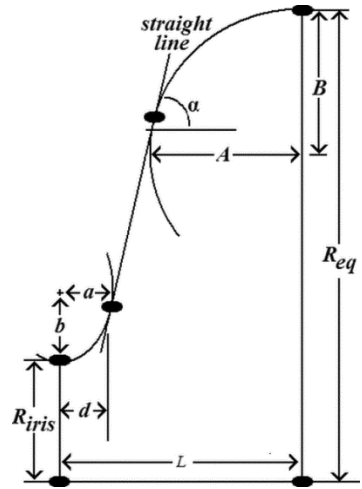

Fig.4

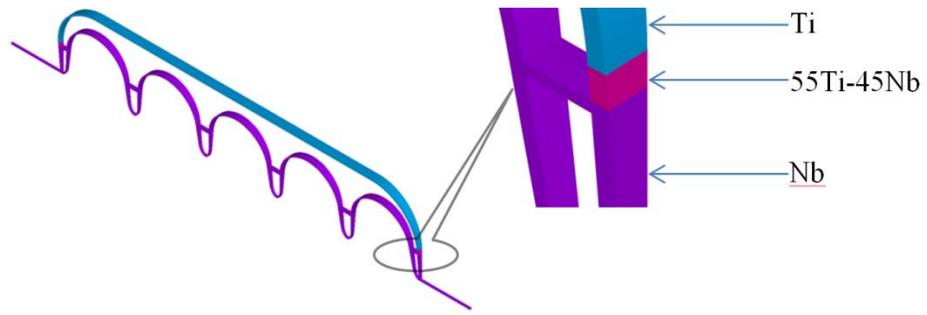

Fig.5

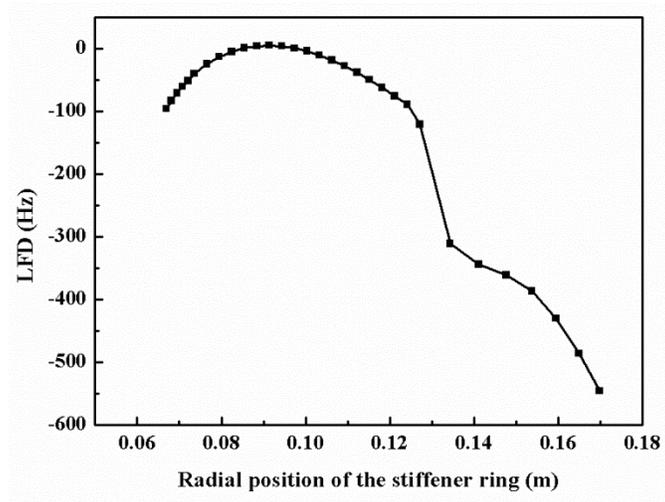

Fig.6

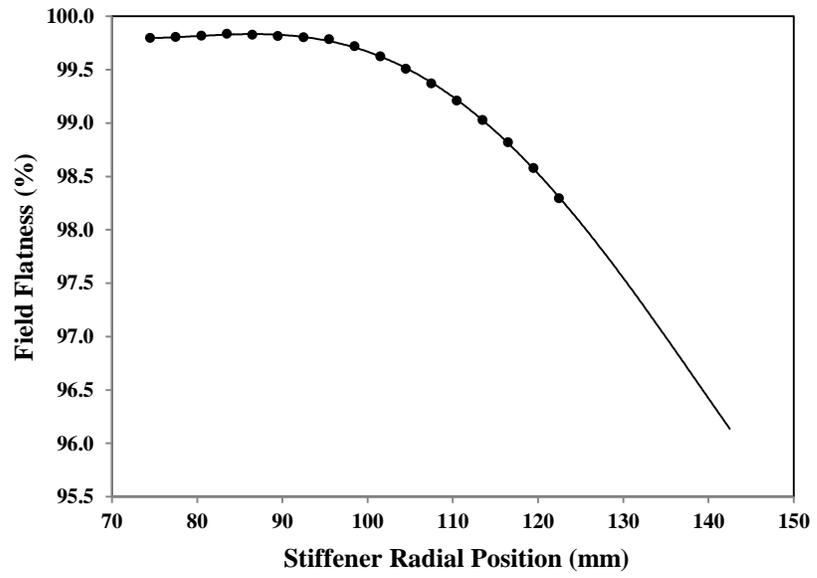

Fig.7

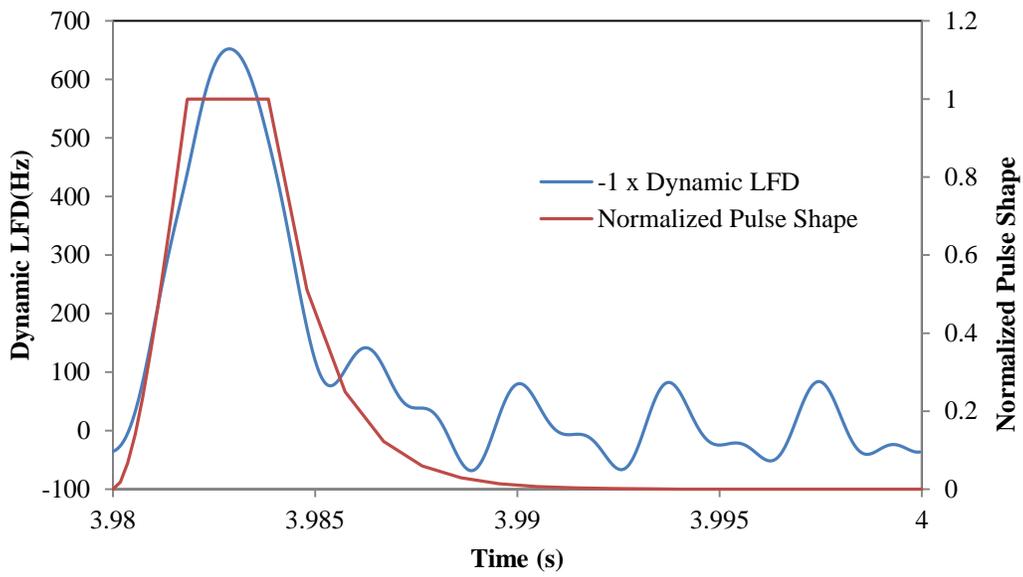

Fig.8

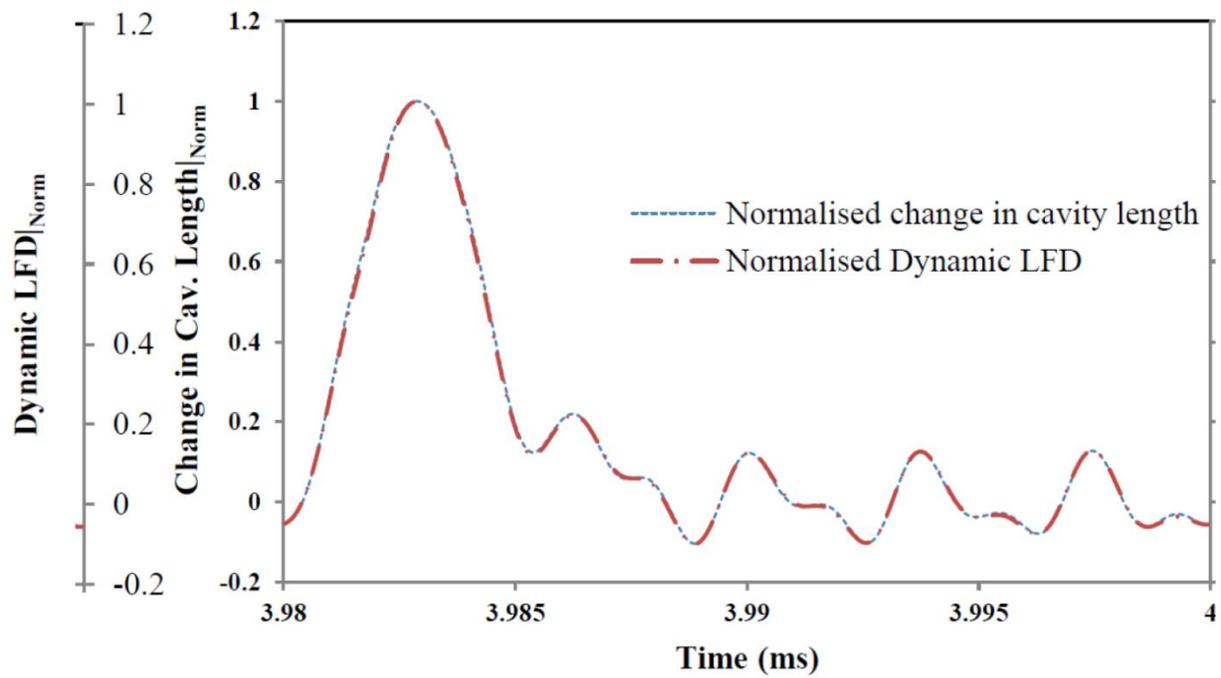

Fig.9

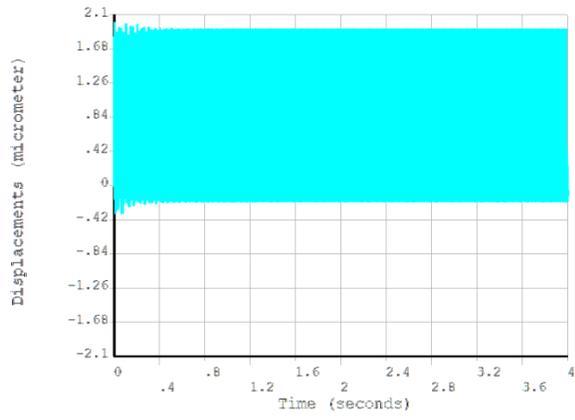 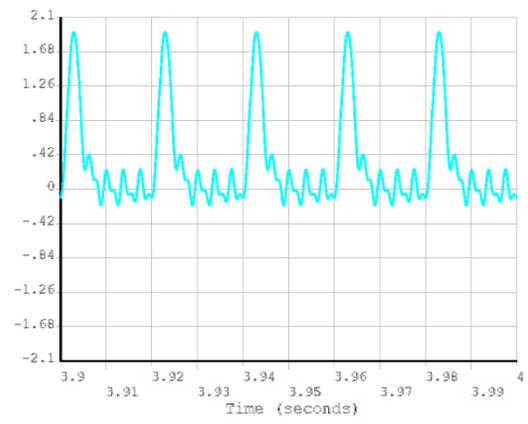

Fig.10

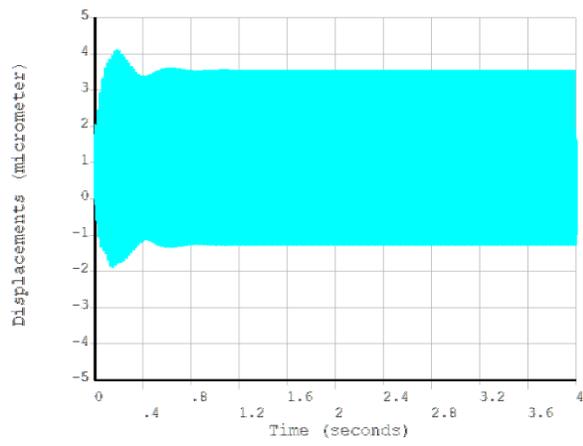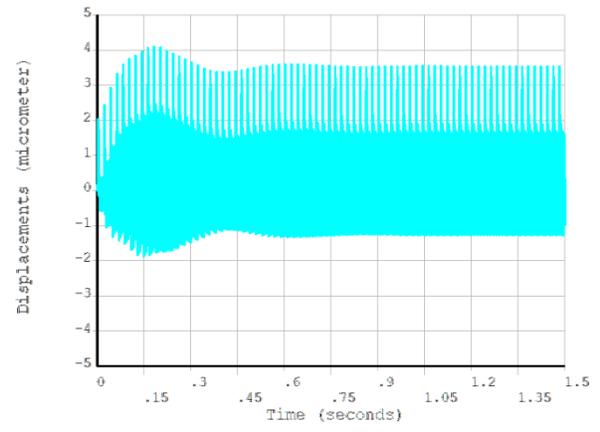

Fig.11

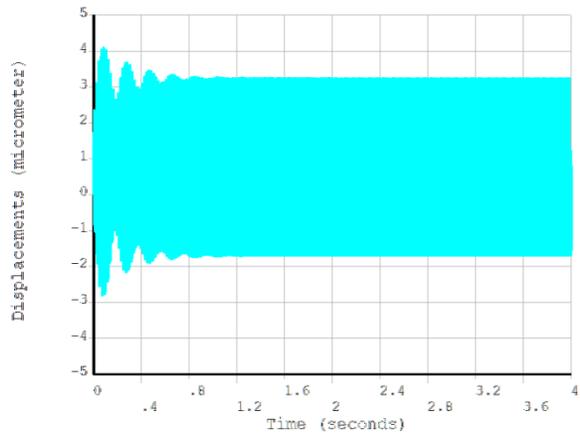 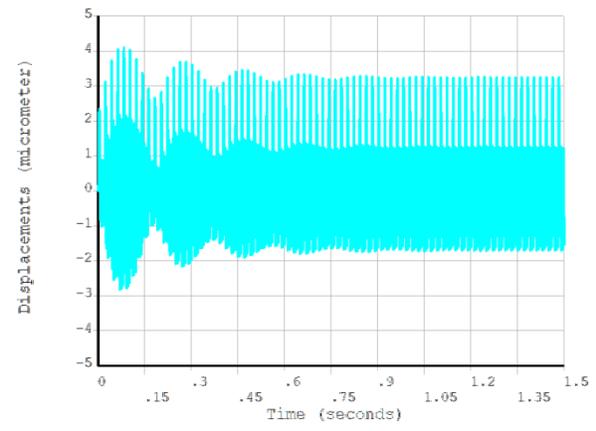

Fig.12

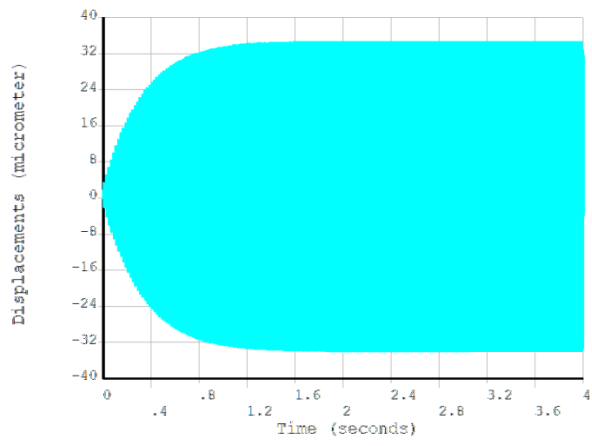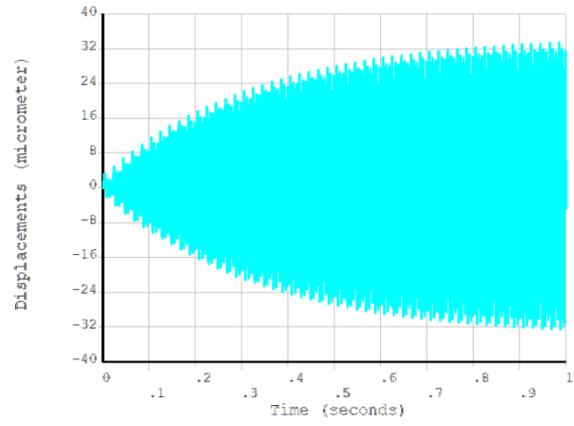

Fig.13

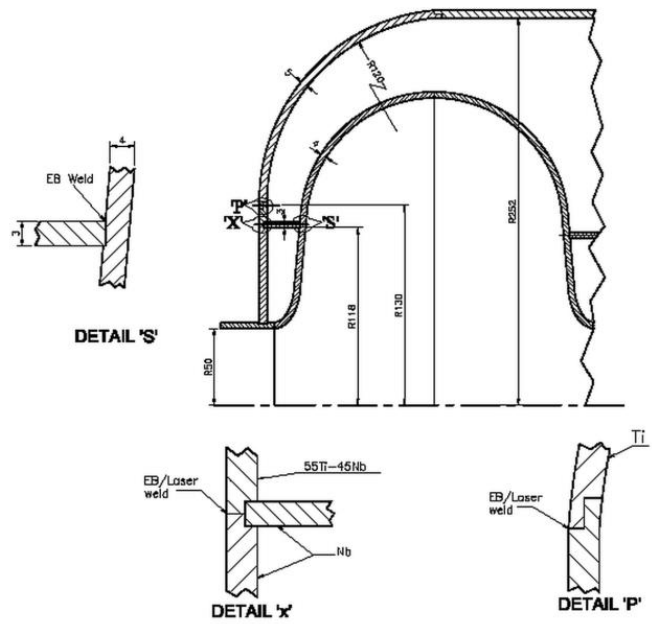

Fig.14